# Hole polaron assisted oxygen ion migration in Li₂MnO₃

Yanhao Dong[1] (dongyh@mit.edu), I-Wei Chen[2], and Ju Li[1,3]

*[1]Department of Nuclear Science and Engineering, Massachusetts Institute of*

*Technology, Cambridge, MA 02139, USA*

*[2]Department of Materials Science and Engineering, University of Pennsylvania,*

*Philadelphia, PA 19104, USA*

*[3]Department of Materials Science and Engineering, Massachusetts Institute of*

*Technology, Cambridge, MA 02139, USA*

We simulated oxygen ion migration in layered-structure $Li_2MnO_3$ (or equivalently $Li_{4/3}Mn_{2/3}O_2$) by first-principles calculations [1-4], which is the parent structure of high-capacity cathodes $(Li_{4/3}Mn_{2/3}O_2)_x \cdot (LiMO_2)_{1-x}$ (M=Ni, Co and Mn) for lithium ion batteries. It has NaCl-type structure with ordered cation sublattice and alternating Li and $Li_{1/3}Mn_{2/3}$ cation slabs (**Fig. 1a**). We considered oxygen ion migration via a vacancy mechanism and studied the effect of hole polaron (by adding one extra hole in the supercell) on the charge state and migration of oxygen ions. As schematically shown by **Fig. 1b**, as the octahedrally coordinated oxygen ion (with two Mn and four Li) exchange its location with a neighboring oxygen vacancy, it would pass by two nearest-neighbor Li, so we also studied the effect of one or two interacting Li vacancies. The results of considered defect model, calculated migration barrier, and Bader charges of (to be) migrating oxygen ion at ground state before/after migration and at saddle point of migration are summarized in **Table 1**. Here, we considered fully ionized oxygen vacancy with +2 formal charge and Li vacancy with −1 formal charge with defect chemistry model; for $O^{2-}$ migration, no extra charge is added to/subtracted from the supercell; for $O^-$ migration, one electron is subtracted from the supercell (i.e., one hole is added to the supercell; and first-principles calculations would determine the location where the added hole would be (de-)localized). We found $O^-$ always has a lower migration barrier (~0.7 eV lower, or about 25% lower) than $O^{2-}$: 2.33 eV for $O^-$ (model A) vs. 3.00 eV for $O^{2-}$ (model B) without neighboring Li vacancy; 2.37 eV for $O^-$ (model C) vs. 3.13 eV for $O^{2-}$ (model D) with one neighboring Li vacancy; 2.32 eV for $O^-$ (model E) vs. 3.09 eV for $O^{2-}$ (model F) with two neighboring Li vacancy. We also found the added extra hole tends to localize on the migrating oxygen ion at the saddle point by forming a small hole polaron, while it tends to delocalize at the ground state. This trend can be seen from Bader charge of the migrating oxygen ion for model B, D and F summarized in **Table 1** and from calculated density of states (DOS) for model B in **Fig. 3**, model D in **Fig. 5** and model F in **Fig. 7** (Fermi level $E_F$ at ground state is at the tail of valence band maximum indicating delocalizing hole state, while it is pinned by gap states at saddle point). So forming hole polaron on migrating oxygen ion indeed lowers its migration barrier. Interesting, for model C and E with one or two neighboring Li vacancy and without extra hole, we found oxygen ion also trends to give out electron and have similar Bader charge as the one in model B, D and F at the saddle point. This further suggests the energy advantage of forming hole polaron at saddle-point migrating oxygen ion, even without the presence of extra hole. We confirmed for model C and E, this hole comes from charge transfer between migrating oxygen ion and neighboring Mn at the saddle point, in a sense that the reaction $O^{2-}+Mn^{4+}\rightarrow O^-+Mn^{3+}$ happens at the saddle point, which is the reversed situation than the ground state ($O^-+Mn^{3+}\rightarrow O^{2-}+Mn^{4+}$, as illustrated by mere existence of $(Li^+)_2(Mn^{4+})(O^{2-})_3$).

We next simulated oxygen ion migration in delithiated $Li_{2-\delta}MnO_3$ with $\delta$=1.19 (see one calculated atomic structure in **Fig. 8**) and again considered oxygen ion migration via exchange with a neighboring oxygen vacancy. Since oxygen redox $O^{2-}\rightarrow O^-$ instead of $Mn^{4+}$ oxidation is responsible for charge storage during delithiation of $Li_2MnO_3$, there would be many holes introduced into the system. So under such a situation, no extra hole is added to the supercell in our calculations. We obtained migration barriers of 0.59-0.76 eV for oxygen ion migration in $Li_{0.81}MnO_3$ and confirmed very small Bader charge of 6.62-6.84 $e$ for the migrating oxygen ion at saddle point, which again highlights the significance of forming hole polaron on migrating oxygen ion. Such small migration barriers enabled by hole localization would allow for long-range lattice diffusion of oxygen ion at room temperature (a migration barrier of ~3 eV in $Li_2MnO_3$ would have prohibited such a possibility), which explains why continuous oxygen loss and gradual voltage decay take place in lithium excess cathodes for lithium ion batteries. More detailed simulations and experiments will later be available to provide better mechanistic understanding and more practice implications for this new observation.

**Computational details**

All first-principles calculations were performed by the Vienna *ab initio* simulation package (VASP) based on density functional theory (DFT) with projector augmented-wave (PAW) method under Perdew-Burke-Ernzerhof (PBE) generalized gradient approximation (GGA). [1-3] PAW potentials including $2s^1$ electron for Li, $3d^5 4s^2$ electrons for Mn and $2s^2 2p^4$ electrons for O were used. DFT+$U$ approach [4] was used to describe the energy of Mn $3d$ with $U$=3.9 eV and $J$=0 eV. We used 400 eV plane-wave cutoff energy and sampled the Brillouin zone using Monhorst-Pack scheme with a 3×3×3 $k$-point mesh to reach a convergence with residue atomic forces less than 0.05 eV/Å. A supercell containing 32 Li, 16 Mn and 48 O was used, and defects were created according to the descriptions in the text. To simulate $Li_{0.81}MnO_3$, we first randomly removed 19 Li to obtain a supercell containing 13 Li, 16 Mn and 48 O, then annealed the structure by first-principles molecular dynamics at 500 K for 1.5 ps followed by step-like cooling at 400 K, 300 K and 200 K (each for 1.5 ps) and finally relaxed at 0 K. One oxygen vacancy was then randomly created to allow for oxygen migration from a neighboring site. Oxygen migration calculations were all conducted using solid-state dimer method [7], after pre-screening using nudged-elastic-band (NEB) method [8].

**References**

[1] G. Kresse, and D. Joubert, Phys. Rev. B **59**, 1758 (1999).

[2] J.P. Perdew, K. Burke, and M. Ernzerhof, Phys. Rev. Lett. **77**, 3865 (1996).

[3] G. Kresse, and J. Furthmuller, Comp. Mater. Sci. **6**, 15 (1996).

[4] S.L. Dudarev, G.A. Botton, S.Y. Savrasov, C.J. Humphreys, and A.P. Sutton, Phys. Rev. B **57**, 1505 (1998).

[5] K. Momma, and F. Izumi, J. Appl. Crystallogr. **44**, 1272 (2011).

[6] W. Tang, E. Sanville, and G. Henkelman, J. Phys.: Condens. Matter **21**, 084204 (2009).

[7] P. Xiao, D. Sheppard, J. Rogal, and G. Henkelman, J. Chem. Phys. **140**, 174104 (2014).

[8] G. Henkelman, B.P. Uberuaga, and H.A. Jonsson, J. Chem. Phys. **113**, 9901 (2000).

**Table 1** Defect model, calculated migration barrier, and Bader charges of migrating oxygen ion at ground state before/after migration and at saddle point of migration.

| Model | Defect configuration | Migration barrier (eV) | Bader charge of migrating O (e) | |
|:---:|:---:|:---:|:---:|:---:|
| | | | Ground state | Saddle point |
| A | $O^{2-}$ exchange with $V_O^{2+}$ | 3.00 | 7.41 | 7.35 |
| B | $O^-$ exchange with $V_O^{2+}$ | 2.33 | 7.40 | 7.05 |
| C | $O^{2-}$ exchange with $V_O^{2+}$, with one $V_{Li}^-$ | 3.13 | 7.32 | 6.96 |
| D | $O^-$ exchange with $V_O^{2+}$, with one $V_{Li}^-$ | 2.37 | 7.25 | 6.92 |
| E | $O^{2-}$ exchange with $V_O^{2+}$, with two $V_{Li}^-$ | 3.09 | 7.23 | 6.73 |
| F | $O^-$ exchange with $V_O^{2+}$, with two $V_{Li}^-$ | 2.32 | 7.14 | 6.77 |

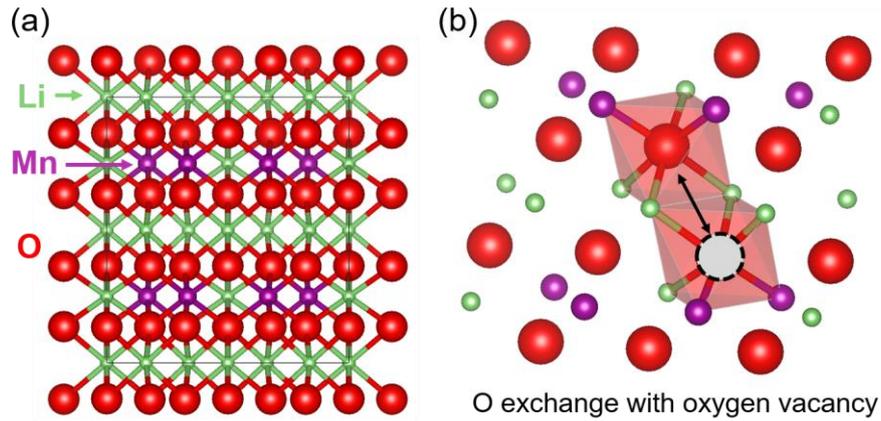

**Figure 1** (a) Atomic structure of Li$_2$MnO$_3$. Li in green, Mn in purple and O in red. (b) Schematic oxygen migration by exchanging one oxygen ion with a neighboring oxygen vacancy. Atomic structure drawn by VESTA [5].

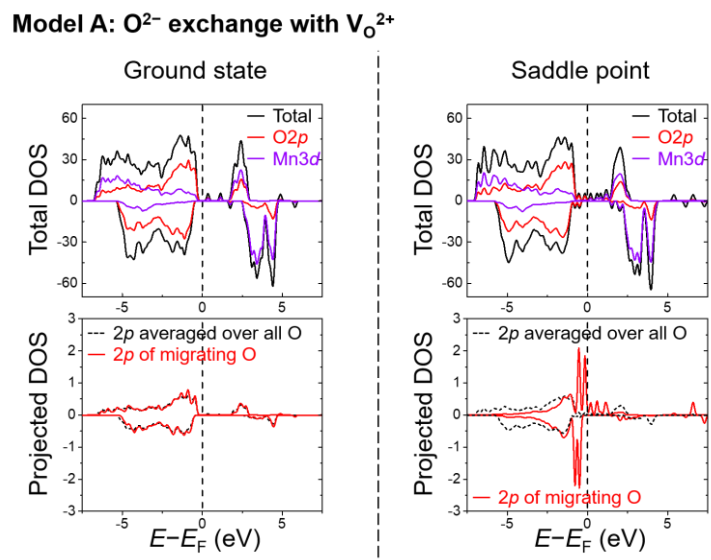

**Figure 2** Calculated total DOS (solid line in black at top panel) and projected DOS for O 2$p$ (solid line in red at top panel), Mn 3$d$ (solid line in purple at top panel), 2$p$ of migrating oxygen ion (solid line in red at bottom panel) and averaged 2$p$ of all oxygen ions for model A. Left panel for ground state and right panel for saddle point.

**Model B: O⁻ exchange with V$_O^{2+}$**

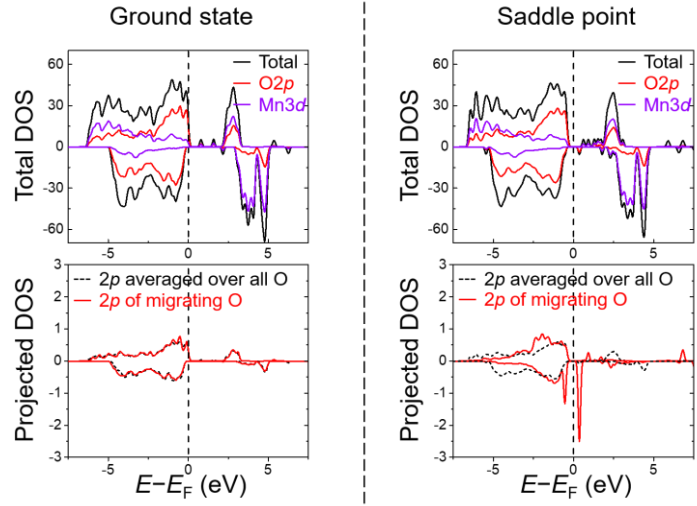

**Figure 3** Calculated total DOS (solid line in black at top panel) and projected DOS for O 2*p* (solid line in red at top panel), Mn 3*d* (solid line in purple at top panel), 2*p* of migrating oxygen ion (solid line in red at bottom panel) and averaged 2*p* of all oxygen ions for model B. Left panel for ground state and right panel for saddle point.

**Model C: O²⁻ exchange with V$_O^{2+}$, with one V$_{Li}^-$**

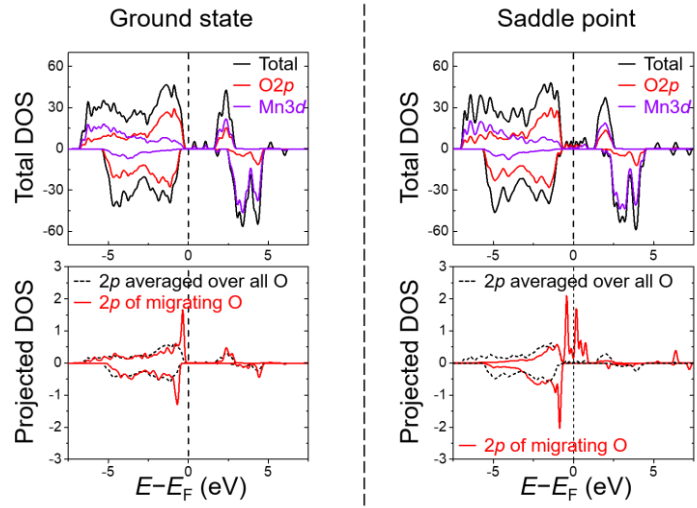

**Figure 4** Calculated total DOS (solid line in black at top panel) and projected DOS for O 2*p* (solid line in red at top panel), Mn 3*d* (solid line in purple at top panel), 2*p* of migrating oxygen ion (solid line in red at bottom panel) and averaged 2*p* of all oxygen ions for model C. Left panel for ground state and right panel for saddle point.

**Model D: O⁻ exchange with V_O²⁺, with one V_Li⁻**

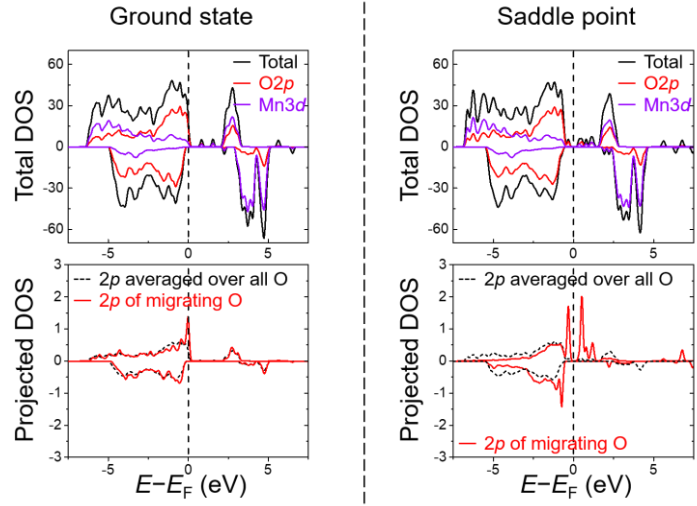

**Figure 5** Calculated total DOS (solid line in black at top panel) and projected DOS for O 2*p* (solid line in red at top panel), Mn 3*d* (solid line in purple at top panel), 2*p* of migrating oxygen ion (solid line in red at bottom panel) and averaged 2*p* of all oxygen ions for model D. Left panel for ground state and right panel for saddle point.

**Model E: O²⁻ exchange with V_O²⁺, with two V_Li⁻**

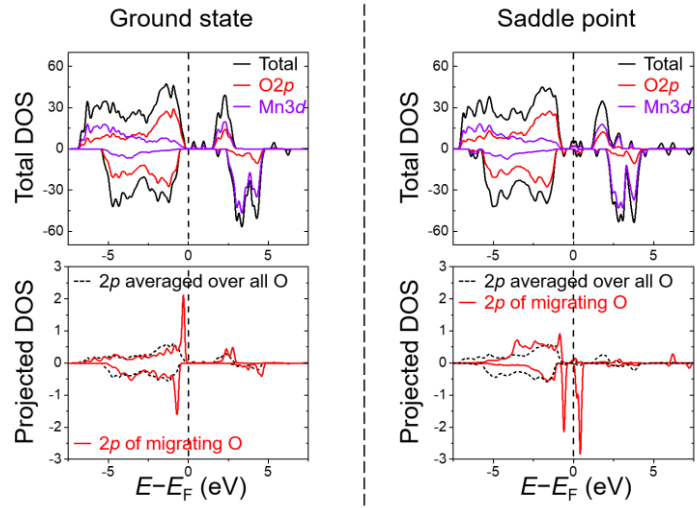

**Figure 6** Calculated total DOS (solid line in black at top panel) and projected DOS for O 2*p* (solid line in red at top panel), Mn 3*d* (solid line in purple at top panel), 2*p* of migrating oxygen ion (solid line in red at bottom panel) and averaged 2*p* of all oxygen ions for model E. Left panel for ground state and right panel for saddle point.

**Model F: O⁻ exchange with V$_O^{2+}$, with two V$_{Li}^-$**

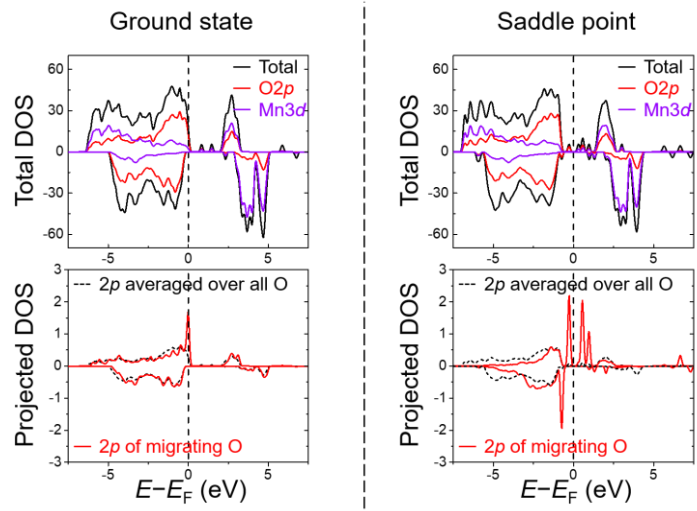

**Figure 7** Calculated total DOS (solid line in black at top panel) and projected DOS for O 2$p$ (solid line in red at top panel), Mn 3$d$ (solid line in purple at top panel), 2$p$ of migrating oxygen ion (solid line in red at bottom panel) and averaged 2$p$ of all oxygen ions for model F. Left panel for ground state and right panel for saddle point.

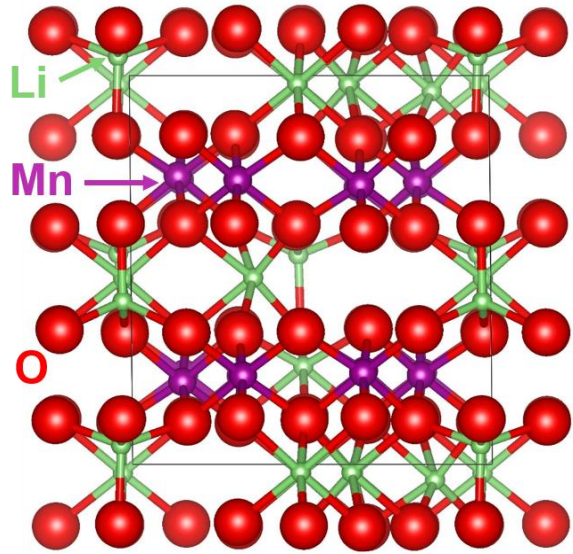

**Figure 8** Atomic structure of one simulated Li$_{0.81}$MnO$_3$.